\documentclass[12pt]{article}

\newcommand{\epo}{e^+_1}
\newcommand{\emo}{e^-_1}

\def\subdef#1{\gdef\globalColor##1{##1}}      
      \subdef{Black}

\newcommand{\plabel}{\label}

\usepackage{epsfig}
\usepackage{amstex}

\begin{document}

\begin{titlepage}
\renewcommand{\thefootnote}{\fnsymbol{footnote}}

\hspace*{\fill} TUW--96--28 \\
\hspace*{\fill} gr-qc/9612012 \\
\begin{center}
\vspace{1cm}

\textbf{\Large Exact Path Integral Quantization of Generic 2-D Dilaton Gravity}\\
\vfill
\renewcommand{\baselinestretch}{1}

\textbf{W. Kummer$^1$\footnotemark[1], H. Liebl$^1$\footnotemark[2]
       and D.V. Vassilevich$^{1,2}$\footnotemark[3]}

\vspace{7ex}

{$^1$Institut f\"ur
    Theoretische Physik \\ Technische Universit\"at Wien \\ Wiedner
    Hauptstr.  8--10, A-1040 Wien \\ Austria}

\vspace{2ex}

{$^2$Department of Theoretical Physics \\
   St. Petersburg University \\ 198904 St. Petersburg \\  Russia}

  \footnotetext[1]{e-mail: \texttt{wkummer@@tph.tuwien.ac.at}}
  \footnotetext[2]{e-mail: \texttt{liebl@@tph16.tuwien.ac.at}}
  \footnotetext[3]{e-mail: \texttt{vasilevich@@phim.niif.spb.su}}


\end{center}
\vfill

\begin{abstract}
Local path integral quantization of generic 2D dilaton gravity
is considered.  Locality means that we assume asymptotic fall
off conditions for all fields.  We demonstrate that in the
absence of `matter' fields to all orders of perturbation theory
and for all 2D dilaton theories the quantum effective action
coincides with the classical one.  
This resolves the apparent contradiction between the well established
results of Dirac quantization and perturbative (path-integral) 
approaches which seemed to yield non-trivial quantum corrections.  
For a particular
case, the Jackiw--Teitelboim model, our result is even extended
to the situation when a matter field is present.
\end{abstract}
\end{titlepage}
\vfill

\section{Introduction}

Semiclassical calculations, i.e.  the quantization of a scalar
field on a classical gravitational background are known to lead
to fascinating results such as the evaporation of black
holes \cite{unruh}.  On the other hand, this additional insight
leads to new fundamental problems, e.g.  the information paradox
or to the question of black hole remnants to name only
the most prominent ones.  A satisfying answer to the final fate
of a singularity can only be extracted from a theory with
quantized matter {\it and} gravity.

In recent years, stimulated by the `dilaton black 
hole' \cite{dilat,kunstatter} numerous studies of quantized gravity 
in the simplified
setting of 2D models were performed (\cite{louis} to \cite{henneaux}). 
Louis-Martinez et al.\ \cite{louis} treated generic 2D dilaton gravity
in the second order formalism
\begin{equation}
 \plabel{lbegin}
{\cal{L}}_{(1)}=\sqrt{-g}
\left(-X\frac{R}{2}-\frac{U(X)}{2}(\nabla X)^2+V(X))\right)
\end{equation}
using a Dirac quantization scheme.
A gauge theoretical formulation for string inspired gravity was developed and
quantized by Cangemi and Jackiw \cite{cangemi-jackiw}. In ref.\ \cite{benedict}
their solutions were shown to be equivalent to the ones of \cite{louis}.
A Dirac approach was recently used to quantize string inspired dilatonic 
gravity \cite{KuRoVa}. In an alternative
approach spherically symmetric gravity was quantized in Ashtekar's
framework by Kastrup \cite{kastrup} and in a geometrodynamical
formulation by Kucha\v{r} \cite{kuchar}.  In particular Strobl
\cite{stroblquant} has treated a large class of 2D gravity theories
within the Poisson-Sigma approach.  
A common feature of all these studies is that due to the particular structure
of the theory the constraints can be solved exactly, yielding a finite
dimensional phase space. Then as a consequence of Dirac quantization
it is found that quantum effects for only a finite number of 
variables are observed. Physically this is in agreement with the fact 
that dilatonic gravity describes no propagating gravitons.
Due to the particular structure of the theory the constraints can be solved
exactly yielding a finite dimensional reduced phase space.
This remarkable property raises hope that in the case of
dilatonic gravity one will be able to get insight into the information
paradox without being forced to deal with the ultraviolet problems of higher dimensional
gravities. Of course, despite of its many appealing features this approach by
itself is insufficient to describe Hawking radiation in quantum
gravity. In the presence of an additional matter field again an infinite
number of modes must  be quantized. In order to tackle that 
problem the results mentioned above first 
of all should be translated into the language of (non 
perturbative) quantum field theory described by the path integral 
as the most adequate method for dealing with infinite dimensional 
quantum systems.

Indeed, one--loop quantum
corrections to the classical action and renormalization group equations 
have been also considered perturbatively \cite{odintsov}.
Matter fields are easily included in this approach. 
In this way, however, even 
pure dilatonic gravity (\ref{lbegin}) was found to exhibit a highly 
non--trivial
renormalization structure, undermining the main motivation for
considering dilatonic gravity as a simple toy model of quantum
black hole physics! 
But even more serious, in our opinion, is the contradiction of these
results with the ones from Hamiltonian approaches as described in the last
paragraph.

Thus a formalism is required which would combine integrability
and simple ultra violet properties of the reduced phase space
quantization with the possibility to include matter and obtain 
the local quantities of the field theoretical quantization.

In the present work we are filling in the gap between the two
approaches mentioned above, removing at the same time the apparent 
contradiction: We demonstrate that in pure
dilatonic gravity (\ref{lbegin}) there are no local quantum
corrections in the effective action for the path integral 
approach as well.
To this end we generalize \cite{kumschwarz,kumhaid} and
perform an exact non-per\-tur\-ba\-tive path integral
quantization of a generic 2D dilaton model containing {\it all} the above models.
We give the explicit form of the generating functional for connected Green
functions. Adding matter fields in general destroys  the 
functional integrability
and suffers therefore from the same weaknesses as the first approach.
However, the particular case of JT gravity \cite{jackiw}  even
in the presence of matter fields allows an exact path integral
quantization\footnote{For the matterless case this model was quantized
exactly already by Henneaux \cite{henneaux}.}.

Main technical features of our approach are the use of the first order
action for Cartan variables in the temporal gauge, corresponding 
to an Eddington Finkelstein (EF) gauge for the 
metric\cite{kumschwarz}. 
Our analysis is local, meaning that
we assume asymptotic fall off  conditions for all fields. This is
enough for the first step of a path integral quantization, which 
of course, in a second step should be adapted to take 
into account global effects familiar from  the reduced phase space approach.

The paper is organized as follows: 
In section \ref{chap621} we show the quantum equivalence of the first order
formulation of gravity including torsion to generic 2D (torsionless)
dilaton gravity. Using the canonical BVF approach in section \ref{chap622}
we obtain the generating functional and find that there are no quantum
corrections to the (matterless) theory. Hence, the effective action coincides
with the original classical one and is therefore only calculated as a
consistency check.
The exact quantization of the JT model {\em including} matter
is presented in section \ref{chap63}.
Critical remarks on necessary boundary conditions and
an outlook on present and future work are contained in section \ref{chap64}.

\section{Exact Path Integral Quantization}
\label{chap62}

\subsection{Equivalence of first and second order forms}
\label{chap621}

Our first goal is to demonstrate the quantum equivalence of the second order form
(\ref{lbegin}) to the first order action 
\begin{equation}
  \plabel{lfirst}
  {\cal{L}}_{(2)}=X^+De^-+ X^-De^++Xd\omega +\epsilon(V(X)+X^+X^-U(X)) ,
\end{equation}
where $De^a=de^a+(\omega \wedge e)^a$ is the torsion two form,
the scalar curvature
$R$ is related to the spin connection $\omega$ by $-\frac{R}{2}=* d\omega$
and $\epsilon$ denotes the volume two form
$\epsilon=\frac{1}{2}\varepsilon_{ab}e^a \wedge e^b=d^2x   \det
e_a^{\mu}=d^2x\,(e)$.  Our conventions are determined by $\eta=diag(1,-1)$
and $\varepsilon ^{ab}$ by $\varepsilon ^{01}=-\varepsilon ^{10}=1$. We
also have to stress that even with Greek indices $\varepsilon^{\mu \nu}$
 is always understood as the antisymmetric symbol and never as the
corresponding tensor. 
The generating functional for the Green functions is given by
\begin{equation}
\plabel{wfirst}
 W=\int ({\cal D}X)({\cal D}X^+)({\cal D}X^-)({\cal D}e^a_{\mu})_{gf}
({\cal D}\omega_{\mu})\exp
\left[i\int_x {\cal{L}}_{(2)} +{\cal{L}}_s \right]\; ,
\end{equation}
where ${\cal{L}}_s$ denotes the Lagrangian containing source terms for the
fields. However, since dilaton gravity does not  any dependence on
$X^{\pm}$ and on $\omega_{\mu}$ we do not introduce the corresponding sources at
this point.
A suitable gauge fixing
is $e_0^-=e_1^+=1$, $e_0^+=0$. It is easy to check that in the following
no division by $e_0^+$ needs to be performed. Also note that
$\det  g = \det  e =1$.
Performing the functional integration with respect to  $\omega_0$ and $\omega_1$ results in
\begin{equation}
\plabel{w2}
 W=\int ({\cal D}X)({\cal D}X^+)({\cal D}X^-)({\cal D}e^a_{\mu})
\delta_{\omega_0}\delta_{\omega_1}\exp
\left[i\int_x \hat{\cal{L}}_{(2)} +{\cal{L}}_s \right] .
\end{equation}
The path integral measure in our gauge is
\begin{equation}
({\cal D}e^a_{\mu})_{gf}=F_{FP}{\cal D}e_1^- \label{degf}
\end{equation}
where $F_{FP}$ is the Faddeev--Popov factor.
We use the abbreviations
\begin{eqnarray}
  \delta_{\omega_0}&=&\delta \left(\partial_1X-X^+e_1^-+X^-e_1^+ 
  \right)\; ,  \\
 \delta_{\omega_1}&=&\delta \left(-\partial_0X+X^+e_0^--X^-e_0^+ 
 \right)\; ,   \\
\hat{\cal{L}}_{(2)}&=&\varepsilon ^{\mu\nu}
\left[X^+\partial_{\mu}e_{\nu}^-+X^-\partial_{\mu}e_{\nu}^++e_{\mu}^+e_{\nu}^-
\left(V(X)+X^+X^-U(X)\right)\right]\; .
\end{eqnarray}
Integration over $X^+$ and $X^-$ finally yields
\begin{equation}
  \plabel{wequivalent}
  W=\int ({\cal D}X)({\cal D}g_{\mu \nu})_{gf} \exp
\left[i\int_x {\cal{L}}_{(1)} +{\cal{L}}_s \right]
\end{equation}
In terms of $g_{\mu\nu}$ our gauge condition becomes the
Eddington--Finkelstein gauge $g_{00}=0$, $g_{01}=1$ with the single
unconstrained component $g_{11}=e_1^-$. The path integral measure
becomes
\begin{equation}
({\cal D}g_{\mu \nu})_{gf}=F_{FP}{\cal D}g_{11} \label{dggf}
\end{equation}
with the same Faddeev--Popov determinant as before (\ref{degf}). As a result
of the commonly used introduction of that determinant in our gauge $F_{FP}$
even turns out to be field independent.
${\cal{L}}_{(1)}$ is exactly given by (\ref{lbegin}). To obtain
(\ref{wequivalent}) we used $(e) \equiv \sqrt{-g}$, $(e)^2g^{\alpha\beta}=
\varepsilon ^{\alpha\gamma}\varepsilon ^{\delta\beta}g_{\gamma\delta}$
and the relation
\begin{equation}
  \plabel{omega}
\tilde{\omega}_{\mu}=\eta_{ab}\frac{\varepsilon^{\alpha\beta}}{(e)}
e_{\mu}^a\partial_{\alpha}
e_{\beta}^b \qquad ,
\end{equation}
the tilde indicating the special case of vanishing torsion
such that in (\ref{lbegin}) 
\begin{equation}
\plabel{Rchap6}
\sqrt{-g}\frac{R}{2}=\varepsilon^{\nu \mu}\varepsilon^{\alpha \beta}
\partial_{\mu}\left(\frac{e_{\nu}^a}{e} \partial_{\alpha} e_{\beta a}\right).
\end{equation} 
Therefore, (\ref{omega}) will only produce the torsionless part of the scalar 
curvature as 
it is given in conventional dilaton theories.
Of course, an additional conformal transformation of the zweibein
would result in additional kinetic terms in the Lagrangian.

Thus the quantum theory of (\ref{lfirst}) is indeed equivalent to the one from
the action (\ref{lbegin}).

\subsection{Canonical BVF Approach}
\label{chap622}

As a first step in our quantization program we use \cite{kumhaid} 
the canonical BVF \cite{brs}
approach in order to obtain the determinants that appear by fixing the gauge
in (\ref{lfirst}).
We will be working in a `temporal' gauge
which corresponds to an Eddington Finkelstein gauge for the metric defined by:
\begin{equation}
  \plabel{gaugefix}
  e_0^+ = \omega_0=0 \quad , \quad e_0^-=1
\end{equation}
Let us choose the canonical coordinates to be
$q_i=(\omega_1,\emo ,\epo )$
and the corresponding canonical momenta 
$p^i=(X,X^+,X^-)$.
The Poisson brackets have the form
\begin{equation}
\plabel{poisson}
\{ q_i(x^1), p^k(y^1)\} =\delta^k_i \delta ,
\quad \delta = \delta (x^1 -y^1)
\end{equation}
For the remaining variables $\bar q_i=(\omega_0,e^-_0,e^+_0)$
the canonical momenta $\bar{p_i}$ vanish (primary constraints).
$\bar{q_i}$ and $\bar{p_i}$ 
 are assumed to obey relations analogous to  (\ref{poisson}).
Dropping a surface term\footnote{A justification for doing this here and
in subsequent steps within our present approach will be given below.},
the ${\bar q}'s$ can be identified
as Lagrange multipliers in the canonical Hamiltonian
\begin{equation}
\plabel{canH}
{\cal H}_{c}=\dot{q}_i p_i - {\cal L}_{(1)}=-{\bar{q}}_i G_i
\end{equation}
thereby enforcing the secondary
constraints
\begin{eqnarray}
G_1 &=& -X^+\emo +X^-\epo +\partial_1 X \nonumber \\
G_2 &=& \partial_1 X^+ +\omega_1 X^+ -
\epo (V(X)+X^+X^-U(X)) \nonumber \\
G_3 &=& \partial_1 X^- -\omega_1 X^- +
\emo (V(X)+X^+X^-U(X)). \plabel{constraints}
\end{eqnarray}
Their Poisson brackets
\begin{eqnarray}
\{ G_1,{G}_2\} &=& -G_2\delta \nonumber \\
\{ G_1,{G}_3\} &=& G_3 \delta \nonumber \\
\{ G_2,{G}_3\} &=& -[(V'(X)+X^+X^-U'(X))G_1+
\nonumber \\
 &\ & +X^-U(X)G_2+X^+U(X)G_3 ]\delta \plabel{algebra} \qquad ,
\end{eqnarray}
show the absence of ternary constraints.
The prime denotes differentiation with respect to $X$.
It should be noted that (\ref{algebra}) is the generalization
of \cite{grosse} where the authors restricted themselves to 
$R^2 + T^2$ gravity \cite{katxx}.
The structure functions ${\cal C}^k_{ij}$ defined through
the equation $\{ G_i,G_j\} ={\cal C}^k_{ij}G_k\delta$ can be used
to construct the BRST charge
\begin{equation}
\Omega =\int d^2 x (b_i \bar p^i +G_i c^i +
\bar b_k{\cal C}^k_{ij}c^ic^j ) \quad ,
\plabel{Omega}
\end{equation}
satisfying the nil-potency condition $\{ \Omega, \Omega \} =0$.
Four types of ghost fields, $b$, $c$, $\bar b$ and $\bar c$ have been introduced,
for which the Poisson antibrackets are
\begin{equation}
\{ \bar b_i , c^j \} = \{ b_i , \bar c^j \} =-\delta^j_i \delta .
\end{equation}

Using these ingredients 
we are able to construct the BRS extension of the canonical
Hamiltonian ${\cal H}_{\mbox{\footnotesize{ext}}}$ and the effective Lagrangian 
${\cal{L}}_{\mbox{\footnotesize{eff}}}$, 
\begin{eqnarray}
{\cal H}_{\mbox{\footnotesize{ext}}}&=& {\cal H}_c 
+c_i \bar q_j {\cal C}^k_{ij}\bar b_k
-b_i \bar b_i . \nonumber \\
{\cal L}_{\mbox{\footnotesize{eff}}}&=&\dot q_i p_i +\dot {\bar q}_i \bar p_i +
\dot b_i \bar c_i + \dot c_i \bar b_i -{\cal H}_{\mbox{\footnotesize{eff}}}
\nonumber \\
{\cal H}_{\mbox{\footnotesize{eff}}}&=&{\cal H}_{ext}-\{ \Psi ,\Omega \} ,
\plabel{Leff}
\end{eqnarray}
in the sense of BVF.
Both ${\cal H}_{\mbox{\footnotesize{ext}}}$ and 
${\cal{L}}_{\mbox{\footnotesize{eff}}}$ are BRS invariant:
\begin{equation}
\{{\cal L}_{\mbox{\footnotesize{eff}}}, \Omega\}=
\{{\cal H}_{\mbox{\footnotesize{ext}}},\Omega\}=0 \quad.
\end{equation}
In (\ref{Leff}) $\Psi$ denotes the gauge fermion which contains the gauge
fixing functions $\chi^i$:
\begin{equation}
\Psi = \bar c_i \chi_i .
\plabel{Psi}
\end{equation}
In the temporal gauge (\ref{gaugefix}), the gauge
fixing functions are derived from 
\begin{equation}
\chi_i =\frac 1\gamma (\bar q_i-a_i) , \quad
a_1=a_3=0,\ a_2=1
\end{equation}
with the limit $\gamma \to 0$ to be taken later. This yields
\begin{equation}
{\cal H}_{\mbox{\footnotesize{eff}}} =
{\cal H}_{\mbox{\footnotesize{ext}}} +\frac 1\gamma \bar p_i
(\bar q_i -a_i)-\frac 1\gamma \bar c_i b_i \quad .
\end{equation}
The path integral can be now presented in the form
\begin{equation}
W=\int ({\cal D}q) ({\cal D}p) ({\cal D}\bar q) ({\cal D}\bar p) ({\cal D} b) ({\cal D} \bar c) ({\cal D} \bar b) ({\cal D}c)
\exp [i \int d^2x ({\cal L}_{eff}+Jp+jq) ].
\plabel{Zbrs}
\end{equation}
As a first step the $b$ and $\bar b$ integrals are performed.
By a change of variables $\bar p\to \gamma \bar p$ and $\bar c\to \gamma \bar c$ 
with a unit Jacobian factor and taking the limit $\gamma = 0$ we obtain
the path integral over the remaining variables with the action
\begin{equation}
{\cal L}'= \dot q_i p_i + \bar q_iG_i -\bar p_i(\bar q_i-a_i)
-c_i(\delta_i^k \partial_0 + \bar q_j {\cal C}^k_{ij})\bar c_k +Jp + jq.
\plabel{Lprime}
\end{equation}
Integration over $\bar p$ and $\bar q$ now is equivalent 
to replacing $\bar q_i$ by $a_i$ everywhere in 
(\ref{Lprime}) as it should be according to (\ref{gaugefix}). 
Finally integration over $c$ and $\bar c$ yields the
determinant
\begin{equation}
F=\det (\delta_i^k \partial_0 +{\cal C}^k_{i2} )=
(\det \partial_0 )^2 \det (\partial_0 + X^+U(X)).
\plabel{FPdet}
\end{equation}
$F$ will be cancelled by the determinants produced by integrating over
$X^{\pm}$ and $X$ below. Note that the operator $(\partial_0 + X^+U(X))$
is unitary equivalent to $\partial_0$: $(\partial_0 + X^+U(X))=e^{-y}
\partial_0 e^y$ , $\partial_0y=X^+U(X)$. Hence $\det (\partial_0 +
X^+U(X))$ is independent of $X^+U(X)$.

After having determined
the contributions of the gauge fixing we will now integrate
over the remaining fields. For this reason we introduce sources
$(j^{\pm},j,J^{\pm},J)$ for the fields $(e^{\mp}_1,\omega_1,X^{\mp},X)$,
respectively.

The generating functional for the Green functions is
\begin{equation}
  \plabel{gen}
  W=\int ({\cal D}X)({\cal D}X^+)({\cal D}X^-)({\cal D}e^+_1)({\cal D}e^-_1)({\cal D}\omega_1)F\exp
\left[i\int_x {\cal{L}}_{(2)} +{\cal{L}}_s \right] ,
\end{equation}
where $F$ denotes the determinant from the previous section,
${\cal L}_{(2)}$ is the gauge fixed part of the Lagrangian (\ref{lfirst}) and
${\cal L}_s$ denotes the contribution of the sources:
\begin{equation}
  \plabel{source}
  {\cal{L}}_s = j^+e_1^- + j^-e_1^+ + j\omega_1 +J^+X^- +J^-X^+ +JX
\end{equation}
Integrating over the zweibein components and the spin connection results in
\begin{equation}
  \plabel{gen2}
  W=\int ({\cal D}X)({\cal D}X^+)({\cal D}X^-) \delta_{(1)}\delta_{(2)}
\delta_{(3)}F \exp \left[i \int J^+X^- +J^-X^+ +JX \right]
\end{equation}
with
\begin{eqnarray}
  \plabel{delta}
  \delta_{(1)}&=&\delta \left(-\partial_0 X^+ +j^+ \right) \\
 \delta_{(2)}&=&\delta \left( -(X^+U(X)+\partial_0) X^- +j^- -V(X) \right) \\
 \delta_{(3)}&=&\delta \left(-\partial_0 X +X^+ +j \right) .
\end{eqnarray}
We must stress again that due to the partial integrations where 
we discarded the surface terms the following results are true 
only locally with a proper definition of what we mean when 
Green functions like $\partial_0^{-1}$, $\partial_0^{-2}$ appear.

The remaining integrations can be performed most conveniently in the order
$X^+, X^-$ and finally $X$ to yield for the generating functional 
of connected Green functions
\begin{equation}
  \plabel{Z}
  Z=-i \ln W =\int JX +J^-\frac{1}{\partial_0}j^+ +
J^+ \frac{1}{\partial_0 +U(X)\frac{1}{\partial_0}j^+}
\left(j^- -V(X)\right),
\end{equation}
where $X$ has to be replaced by
\begin{equation}
  \plabel{X}
  X=\frac{1}{\partial_0^2}j^+ +\frac{1}{\partial_0}j \qquad .
\end{equation}
It should be stressed that the determinant $F$ is precisely canceled by the
determinants appearing during these last three integrations.
Eq.(\ref{Z}) gives the exact non-perturbative generating functional
for connected Green functions and it does not contain any 
divergences, because it clearly describes tree--graphs only. 
Hence no quantum effects remain. 

Now we turn to the ill defined expressions $(\partial_0)^{-1}$.
As shown in \cite{kumschwarz} a proper (vanishing) asymptotic behavior results
from a regularization 
\begin{eqnarray}
\plabel{ifcut}
\frac{1}{{\partial_0}^2} &=& 
\lim_{\mu^2 \to 0} \frac{1}{{\partial_0}^2+\mu^2}  \\
\frac{1}{\partial_0} &=& \partial_0 \left( \frac{1}{\partial_0^2} \right) ,
\end{eqnarray}
where $\mu^2$ insures a proper infrared cutoff.
Since each partial integration above involved either $X$, $X^+$ or $X^-$
which in turn all exhibit at least a $\partial_0^{-1}$ behavior,
all our previous steps are justified.

\subsection{Effective Action $\Gamma$}
\label{chap623}

It is illustrative to calculate the effective action $\Gamma$
which generates one particle irreducible vertex functions. $\Gamma$ is defined
through a Legendre transformation of the generating 
functional for connected Green functions
\begin{equation}
\Gamma (\bar X^\pm ,\bar X,\bar e_1^\pm ,\bar\omega_1 )= Z (J,j)-
\int (J^+ \bar X^-+J^- \bar X^+ +J\bar X+j\bar\omega_1 +j^+ 
\bar e^-_1 +j^- \bar e^+_1)\; .
\plabel{defG}
\end{equation}
Here $\bar X,\ \bar e,\ \bar\omega$ denote  the 
corresponding `mean fields' in the presence of sources, e.g.\ 
\begin{equation}
{\bar X^\pm} = \frac {\delta Z}{\delta J^\mp}\; .
\plabel{defX}
\end{equation}
In the following we drop the bars because no confusion can arise 
with the variables in the previous subsection. 
Only three such equations are needed:
\begin{eqnarray}
X &=& \frac 1{\partial_0} \left ( \frac 1{\partial_0} j^+ +j \right ) \nonumber \\
X^+ &=& \frac 1{\partial_0} j^+ \nonumber \\
X^- &=& \frac 1{\partial_0 +U(X)X^+} (j^--V(X))
\plabel{xandj}
\end{eqnarray}
By substituting (\ref{xandj}) in (\ref{defG}) we obtain
\begin{equation}
\plabel{Gamma}
\Gamma = \int (-e^+_1 (\partial_0 + U(X)X^+)X^-
-e^+_1 V(X) - e^-_1 \partial_0 X^+ - \omega_1
\partial_0 X +\omega_1 X^+ )
\end{equation}
Up to surface terms, this is exactly the classical action (\ref{lfirst})
in the temporal gauge (\ref{gaugefix}).
As demonstrated already above for the connected Green functions, there are no 
quantum corrections to the theory, 
(\ref{Gamma}) is certainly the expected result and can hence be considered as
a consistency check.
 
Our result means that up to possible global effects
the quantum effective action should coincide with the classical
one on shell in any parameterization for either the first or
second order formalism. Of course,  in a framework of 
perturbative expansions for certain parameterizations in 
configuration space, combined with unsuitable gauge fixings in the 
case of gauge theories,  artificial divergences can appear,
a familiar example being the non-renormalizable ones  
of individual graphs in the unitary gauge of Yang Mills theories.
In dimensional regularization with the background field gauge in \cite{odintsov}
divergent counterterms were produced which are indeed pure artifacts of this 
approach.
The essential indication for that is their vanishing on shell.
To illustrate this point let us consider a generic theory
described by a classical action $S(\Phi )$. The one--loop effective
action is given by 
\begin{equation}
Z^{\rm 1-loop}_\Phi \propto
\log \det \left ( \frac {\delta S }{ \delta \Phi^A 
\delta \Phi^B} \right )
\plabel{ZPhi}
\end{equation}
In another parameterization\footnote{corresponding to a canonical 
transformation in phase space} $\phi (\Phi)$,  we have
\begin{eqnarray}
Z^{\rm 1-loop}_\phi \propto
\log \det \left ( \frac {\delta S}{\delta \phi^a
\delta \phi^b} \right )\nonumber \\ =
\log \det \left ( \frac {\delta S}{\delta \Phi^A 
\delta \Phi^B} \frac {\delta \Phi^B}{\delta \phi^b}
\frac {\delta \Phi^A}{\delta \phi^a}+ 
\frac {\delta S}{\delta \Phi^C} \frac {\delta^2 \Phi^c}{\delta
\phi^a \delta \phi^b}
\right )
\plabel{Zphi}
\end{eqnarray}
Since the factors $\delta \Phi / \delta \phi$ can be absorbed
into a redefined path integral measure, on shell (\ref{ZPhi}) coincides with
(\ref{Zphi}). This is not true off shell. Even if (\ref{ZPhi}) is
finite, (\ref{Zphi}) may contain divergences. Our result proves that 
the latter may be cancelled 
by a suitable all-order counterterm as in \cite{kumschwarz}
for all 2D dilaton theories. 

\section{JT Model with Matter}
\label{chap63}

Up to now we have quantized theories of pure gravity ignoring the effects of
any matter contribution. In this section we shall minimally couple a scalar 
field to our gravitational action. The standard procedure for path integral
quantization of the scalar field (in the absence of sources for 
that field) amounts to determine
\begin{equation}
\plabel{box}
det \Box = det \frac{1}{\sqrt{-g}}\partial^{\mu}\sqrt{-g}g_{\mu \nu}
\partial^{\nu}
\end{equation}
which in turn (see e.g. \cite{abdalla}) generates the effective nonlocal
action
\begin{equation}
\plabel{polyakov}
{\cal L}_P=\sqrt{-g}R\frac{1}{\Box}R \quad ,
\end{equation}
also called {\it Polyakov} action. $R$ denotes the scalar curvature for 
vanishing torsion (\ref{Rchap6}), which in our temporal gauge 
(\ref{gaugefix}) becomes
\begin{equation}
\plabel{Rgaugefix}
R=\frac{2}{e_1^+} 
\left( \partial_0 \partial_0e_1^- +\partial_0%
\left( \frac{e_1^-}{e_1^+} \partial_0e_1^+\right)-
\partial_1\left( \frac{1}{e_1^+} \partial_0e_1^+\right)\right) .
\end{equation}
The d'Alembertian  as defined in (\ref{box}) has to be expressed in terms of
the zweibeins which in our gauge leads to 
\begin{equation}
\plabel{boxgaugefix}
\Box=\frac{1}{e_1^+} 
\left(\left(\partial_1-e_1^- \partial_0\right)e_1^+\partial_0+
\partial_0e_1^+\partial_1-
e_1^-\partial_0e_1^+\partial_0\right) .
\end{equation}

We see that (\ref{polyakov}) is not linear in the zweibein 
anymore. 
Therefore, in general, adding the Polyakov term to our first 
order action (\ref{lfirst})
will destroy the procedure of section \ref{chap62}. However, there exists a
particular case, the well-known Jackiw Teitelboim (JT) model \cite{jackiw}  
for
which we are able to perform a complete integration of the gravitational
{\bf and} the matter action. The JT model is obtained by setting
$U(X)=0$ and $V(X)=\alpha X$, with $\alpha$ being a constant. 
Performing the integration, all the steps up to 
(\ref{gen}) and (\ref{source}) remain the same because the 
Polyakov term is not affected. 
Due to the complicated dependence of this term on the zweibein one
cannot analytically integrate these fields. Instead we now 
perform the $X$, $X^+$
and $X^-$ integration first to arrive (starting from (\ref{gen})) at 
\begin{equation}
  \plabel{genchap6}
  W=\int ({\cal D}e^+_1)({\cal D}e^-_1)({\cal D}\omega_1)
\delta_{(X)}\delta_{(X^+)}
\delta_{(X^-)}F \exp i \int {\cal L}_{eff}
\end{equation}
with
\begin{eqnarray}
\plabel{chap6Leff}
{\cal L}_{eff}&=&j\omega_1 + j^+ e^-_1 + j^- e^+_1 +{\cal L}_P \\
  \plabel{deltaX}
  \delta_{(X)}&=&\delta \left(\partial_0 \omega_1 +J-\alpha e^+_1 \right) \\
 \delta_{(X^+)}&=&\delta \left(\omega_1 +J^- +\partial_0 e^-_1 \right) \\
\plabel{deltaX3}
 \delta_{(X^-)}&=&\delta \left(\partial_0 e^+_1  +J^+ \right) 
\end{eqnarray}
in a suggestive notation.
The remaining integrations can be done with the use of
(\ref{deltaX}) to (\ref{deltaX3}).
A factor of $\left( \det \partial_0\right)^{-3}$ 
appearing during these calculations again is cancelled by the term $F$ 
under the integral.
As a final result we arrive at 
\begin{equation}
  \plabel{ZJT}
Z=-i \ln W = \int {\cal L}_{eff} ,
\end{equation}
${\cal L}_{eff}$ being defined by (\ref{chap6Leff}).  
The zweibein and connection are understood to be expressed as 
\begin{eqnarray}
 e^-_1 &=&\frac{1}{(\partial_0)^3} 
       \left(\partial_0 J+\alpha J^+ - \partial_0\partial_0J^- \right) \\
 e^+_1 &=&-\frac{1}{\partial_0}J^+\\
\omega_1  &=&-\frac{1}{\partial_0}J -\frac{\alpha}{(\partial_0)^2}J^+  \quad . 
\end{eqnarray}
Thus in  (\ref{ZJT}) we succeeded to give the exact non-perturbative
generating functional for connected Green functions,
quantized `locally', even in the presence of matter.
Our result means that in the absence of external matter sources the quantum
JT model with quantum  matter is locally equivalent to the classical JT
model with the Polyakov term, i.e. the `semiclassical' approximation
becomes exact.
It has to be stressed that with the above method the JT model can be treated 
with any kind of matter (also including fermions\cite{kumxx}) as long as there is no coupling to $X$ or $X^{\pm}$.
This is clear from the fact that the path integral quantization 
of the matter field
results in a purely geometrical term and the subsequent integrations of the 
zweibein and connection can always be achieved with the help of 
(\ref{deltaX}) to (\ref{deltaX3}).

\section{Remarks and Outlook}
\plabel{chap64}
We conclude that the path integral approach is very efficient
for the quantization of dilatonic gravity. We were able to quantize the wide
class of models (\ref{lbegin}). This is the counterpart of the reduced phase space
quantization as considered by Louis-Martinez et al \cite{louis}.
However, it must  be noted  that they actually quantized a model
whose kinetic term for the dilaton field was removed by means of a conformal
transformation. However,
as shown in detail in \cite{kat95}
such transformations drastically change the global
structure of the theory already at the classical level,
and it is therefore by no means clear
how the quantum theory is affected. As we demonstrated here, {\it local}
quantum effects are not affected by that conformal transformation.
There are sources of conformal non--invariance (see e.g.
\cite{ccmann,LVA}) which may change global effects. 
 Therefore this problem deserves further investigation.

Although we succeeded to quantize `all' 2D dilaton gravity models
in a (non-perturbative) path integral formulation 
and to remove in this way the existing apparent contradiction to previous
Hamiltonian approaches,
some self-critical remarks are in order.
We have systematically discarded surface integrals produced by partial integrations.
This approach was implemented by applying an infrared cutoff. However,
since the zweibein will rather exhibit
a certain nonvanishing behavior at `infinity' (e.g. $e_1^+ \to 1$ for
flat space-time) than go to zero, this procedure may be questionable.
The nontrivial quantum effects from nontrivial global compactifications 
including kink solutions \cite{sch93}, now well known for the Hamiltonian 
approach, are eliminated under our assumptions. A more
elaborate treatment, namely keeping track of all boundary terms and the
application of proper, preferably gauge independent boundary conditions
\cite{vasiboundary} is currently under progress \cite{quantboundary}.
This --- in our opinion --- would provide the only way for a 
proper path integral treatment, including global effects as well. 

\section*{Acknowledgement}

This work has been supported by Fonds zur F\"orderung der
wissenschaftlichen For\-schung (FWF) Project No.\ P 10221--PHY.  One
of the authors (D.V.) thanks GRACENAS for financial support.

\vfil

\end{document}